\documentclass[aps,pre,floatfix,twocolumn,showpacs]{revtex4}

\usepackage{graphicx}
\usepackage{epsfig}

\begin{document}

\def\figureDiagram{
\begin{figure*}[t]
 \centerline{\includegraphics[width=\textwidth]{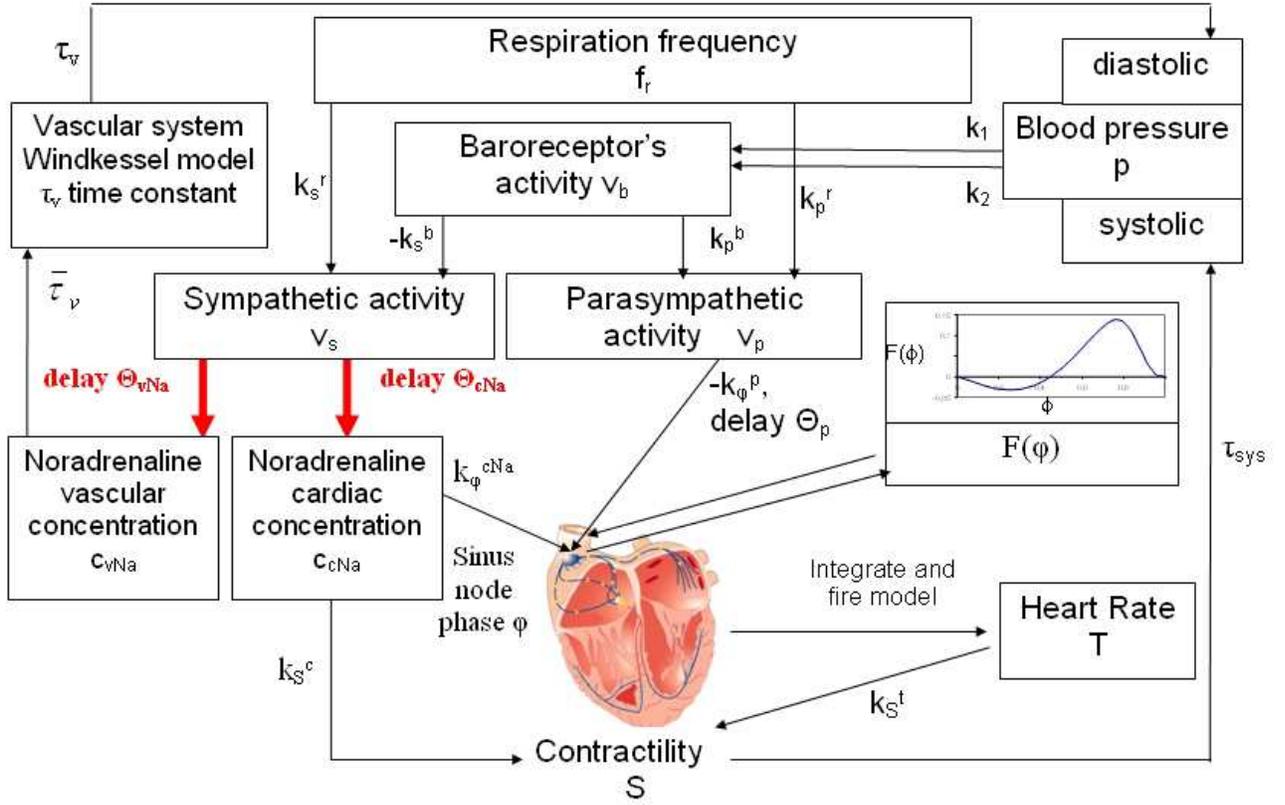}}{
 \caption{\label{SH_scheme} The scheme of SH model of the baroreceptor control loop. Arrows depict relations between variables.}
  }
\end{figure*}
}

\def\figureTau{
\begin{figure}[thb]
\centerline{\includegraphics[width=0.5\textwidth]{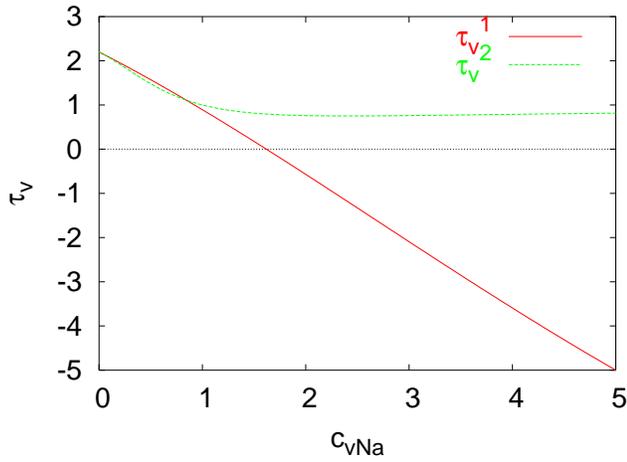}} {
 \caption{\label{figtauv} \small{The dependence $\tau_{v}$ on $c_{vNa}$.} $\tau_v^1=\tau_{v}(c_{vNa})$ for $\widehat{c}_{vNa}=10.0$,$\tau_v^2=\tau_{v}(c_{vNa})$ for $\widehat{c}_{vNa}=1.0$}
  }
\end{figure}
}

\def\figureSimulation{
\begin{figure*}
\begin{center}
$\begin{array}{c@{\hspace{0.2in}}c}
\multicolumn{1}{l}{\mbox{\bf (a)}} &
    \multicolumn{1}{l}{\mbox{\bf (b)}} \\ [0 cm]                	\includegraphics[width=0.45\textwidth]{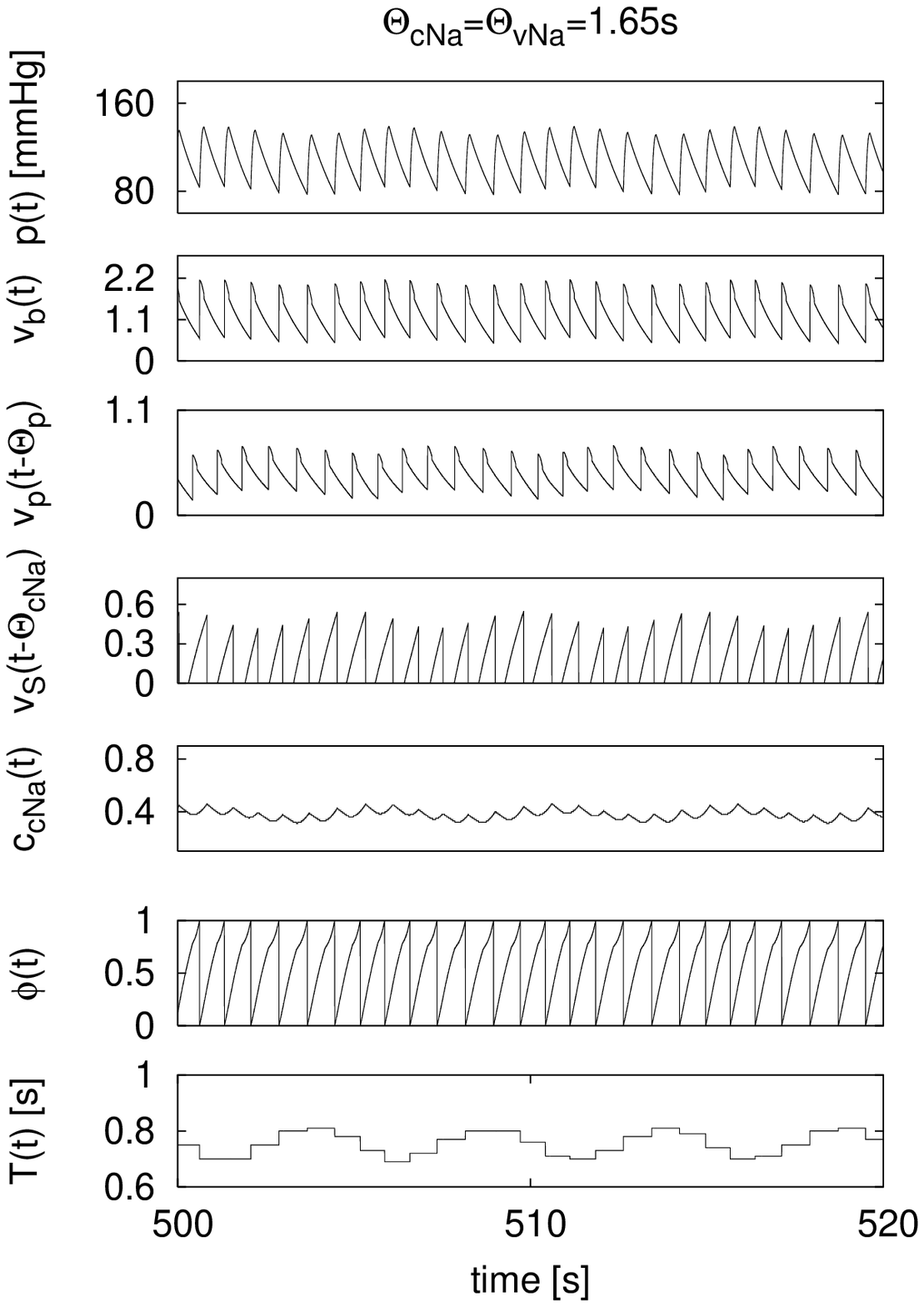}&
\includegraphics[width=0.45\textwidth]{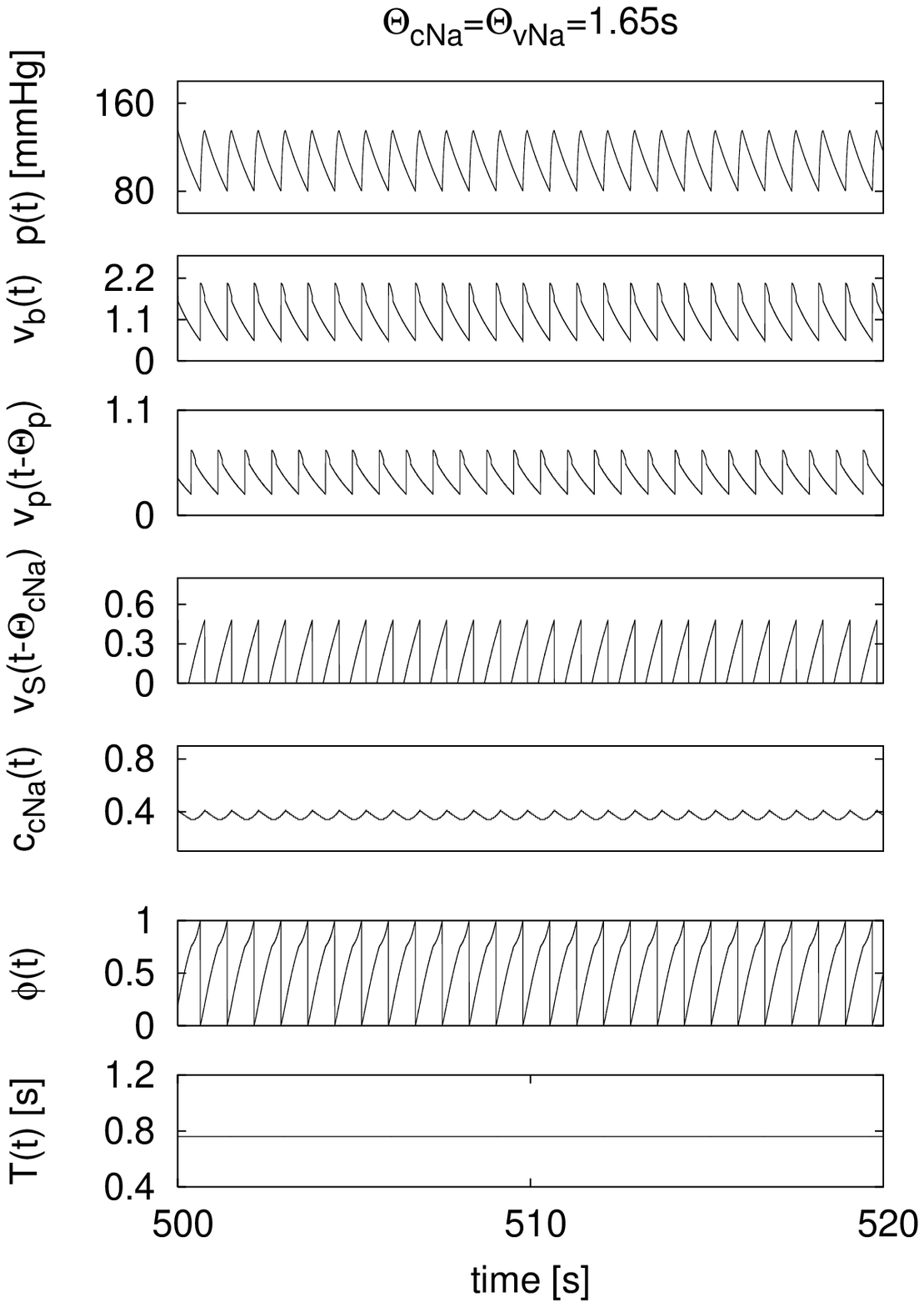} \\ [0.4cm]
\end{array}$
\end{center}
\caption{\label{ts_norand} \small{Time series of the SH model variables. Time delays are fixed $\Theta_{cNa}$=$\Theta_{vNa}$=1.65s. \textbf{(a)}. The influence of respiration can be seen in modulations of waveforms (RSA). \textbf{(b)}. Time series when the respiratory modulation of naural activities is neglected}}
\end{figure*}
}

\def\figureBifcrega{
\begin{figure}
    \includegraphics[width=0.45\textwidth]{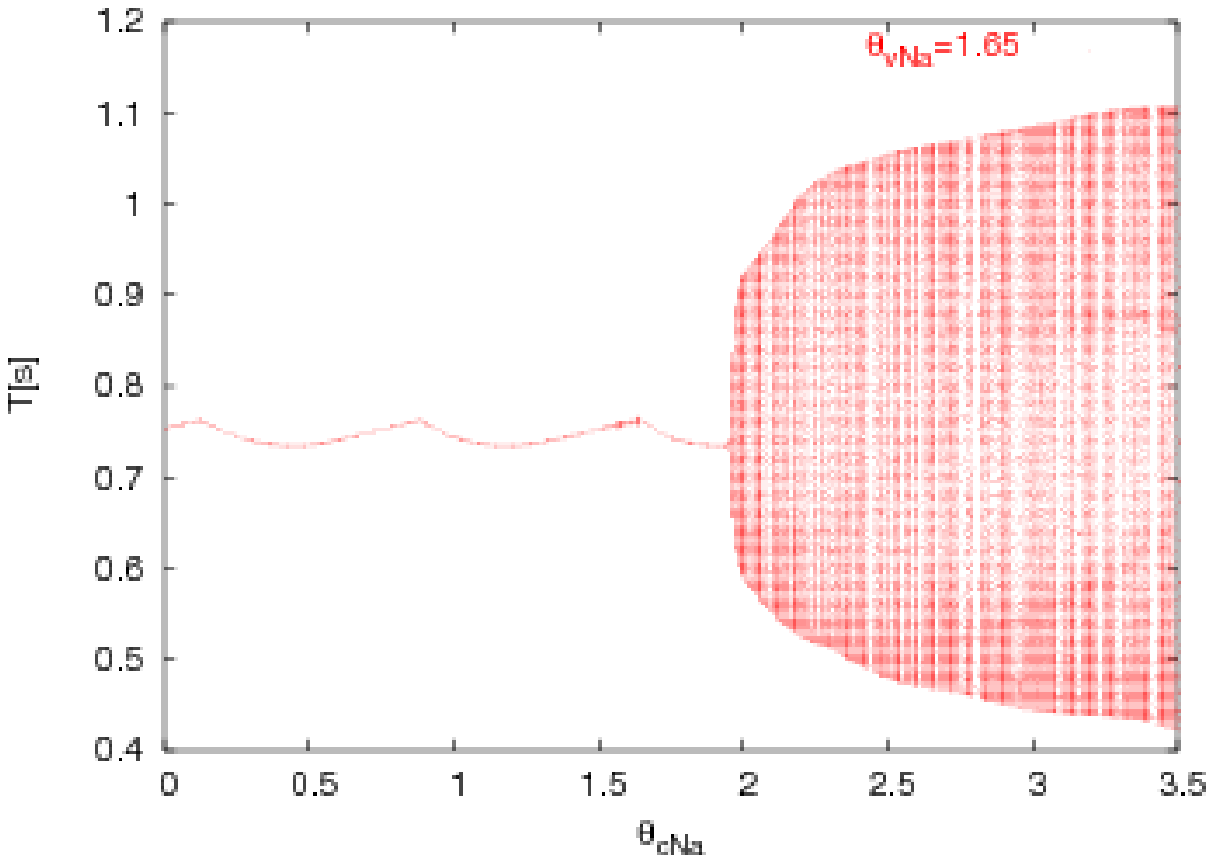}
    \caption{The delay in transmitting signals from the brain to the heart $\theta_{cNa}$ is increasing, while the delay between the brain and the blood vessels is constant  $\theta_{vNa} = 1.65 s$}
  \label{figbifc165}
\end{figure}  
}

\def\figureBifcregb{
\begin{figure}
    \includegraphics[width=0.45\textwidth]{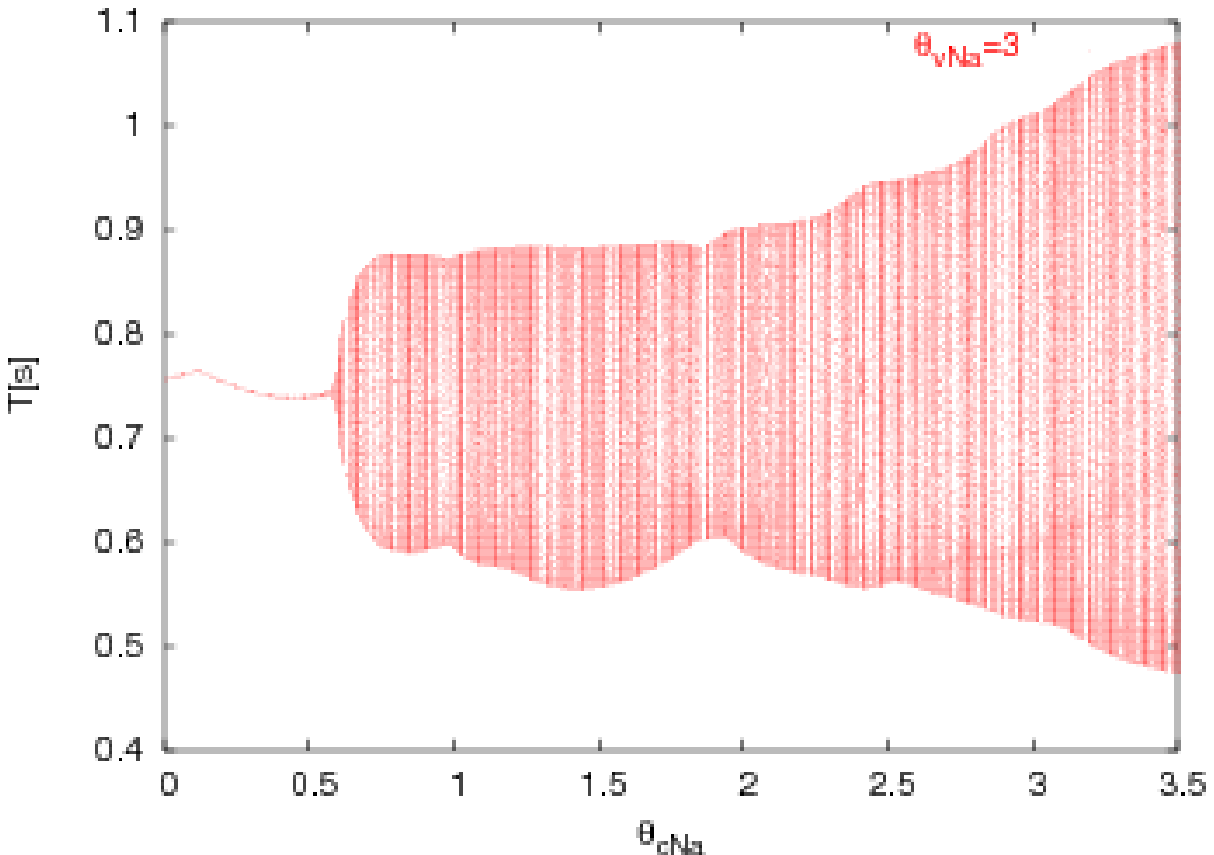}
    \caption{The delay in transmitting signals from the brain to the heart $\theta_{cNa}$ is increasing, while the delay between the brain and the blood vessels is constant  $\theta_{vNa} = 3 s$}
  \label{figbifc3}
\end{figure}  
}

\def\figureSimulationa{
\begin{figure}
\includegraphics[width=0.45\textwidth]{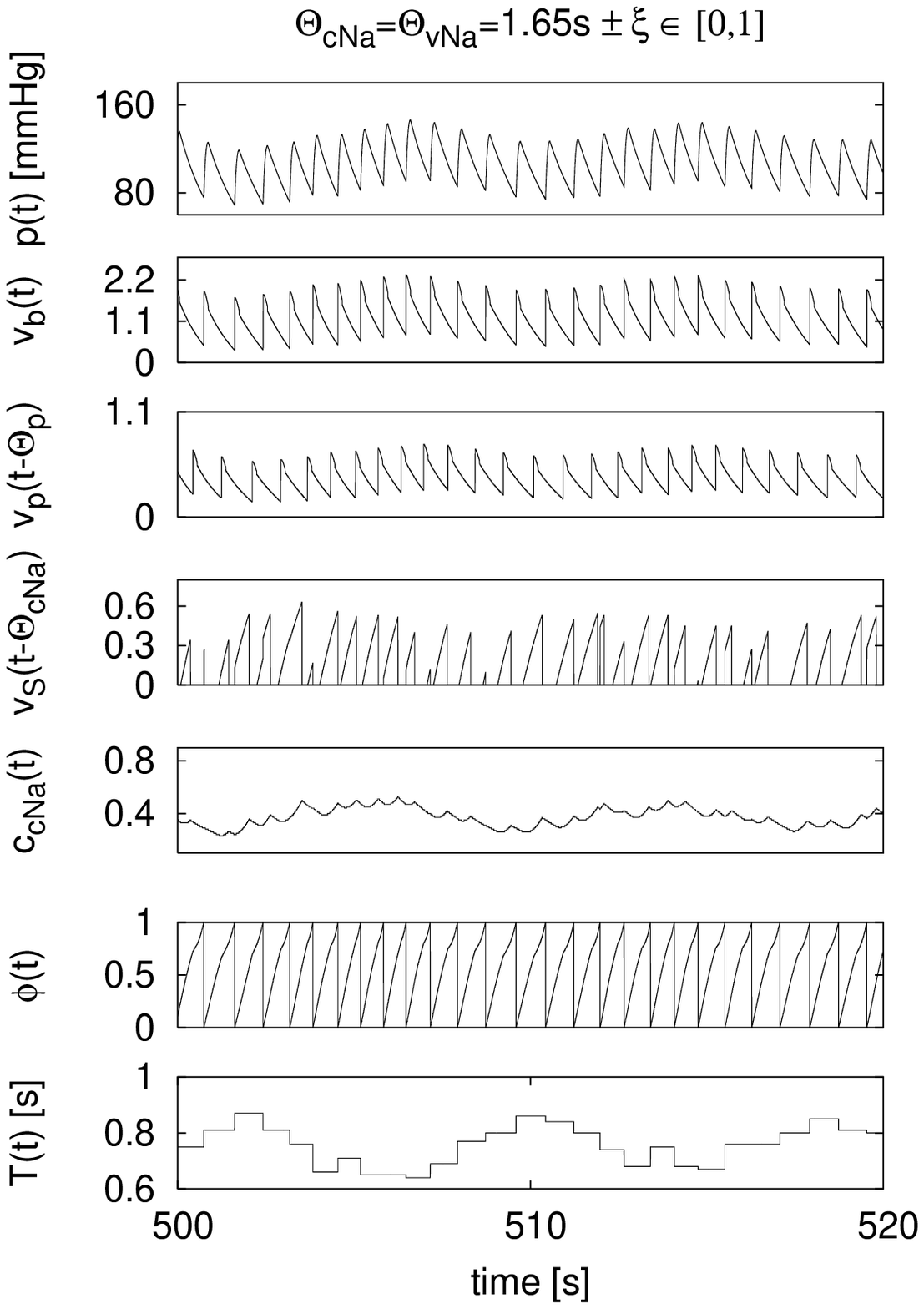}
\caption{\label{sima}Time series obtained by simulating the model with stochastic delays - regular heart rate case.}
\end{figure}
}

\def\figureSimulationb{
\begin{figure*}
\begin{center}
$\begin{array}{c@{\hspace{0.2in}}c}
\multicolumn{1}{l}{\mbox{\bf (a)}} &
    \multicolumn{1}{l}{\mbox{\bf (b)}} \\ [0 cm]
\includegraphics[width=0.45\textwidth]{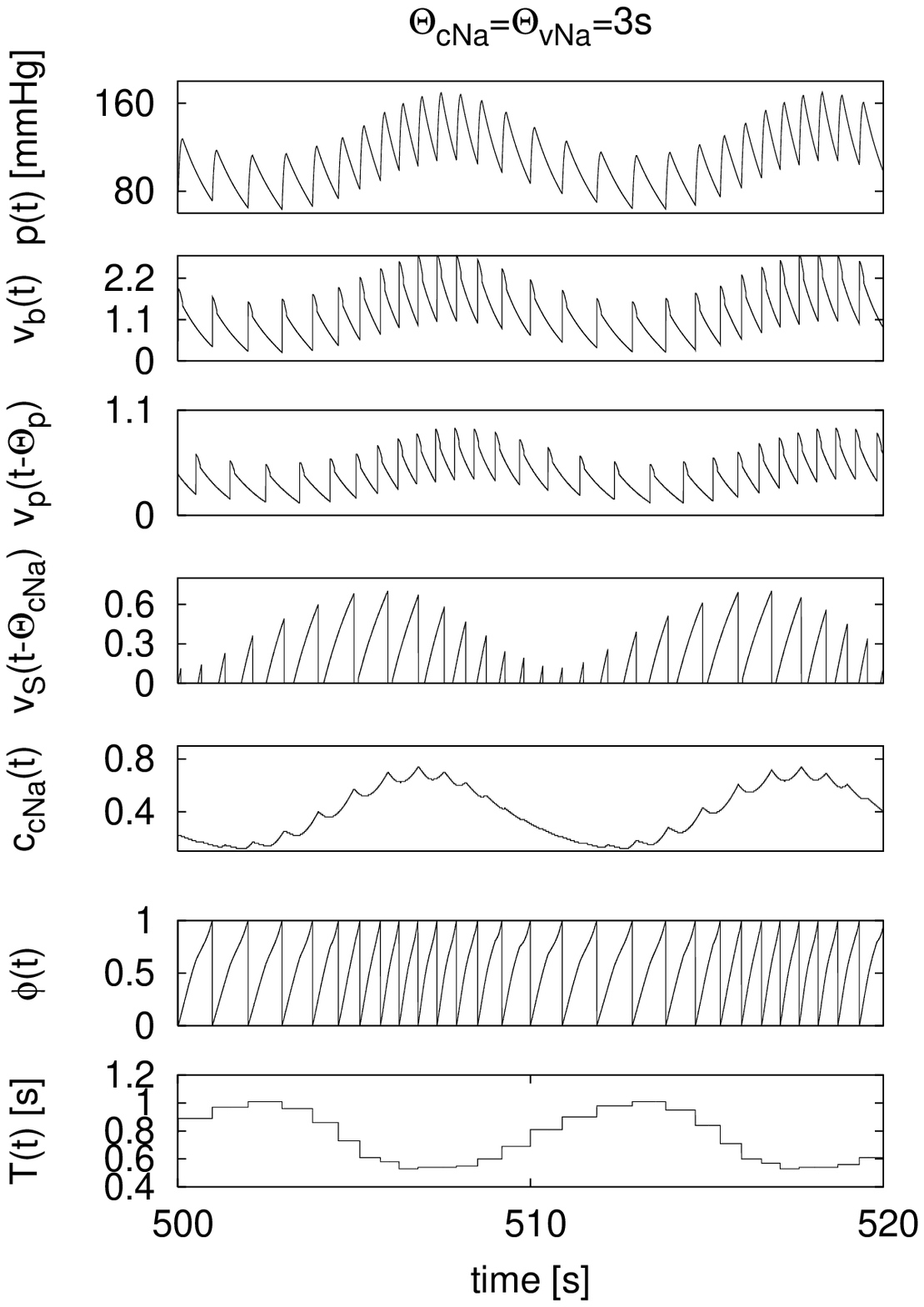}& \includegraphics[width=0.45\textwidth]{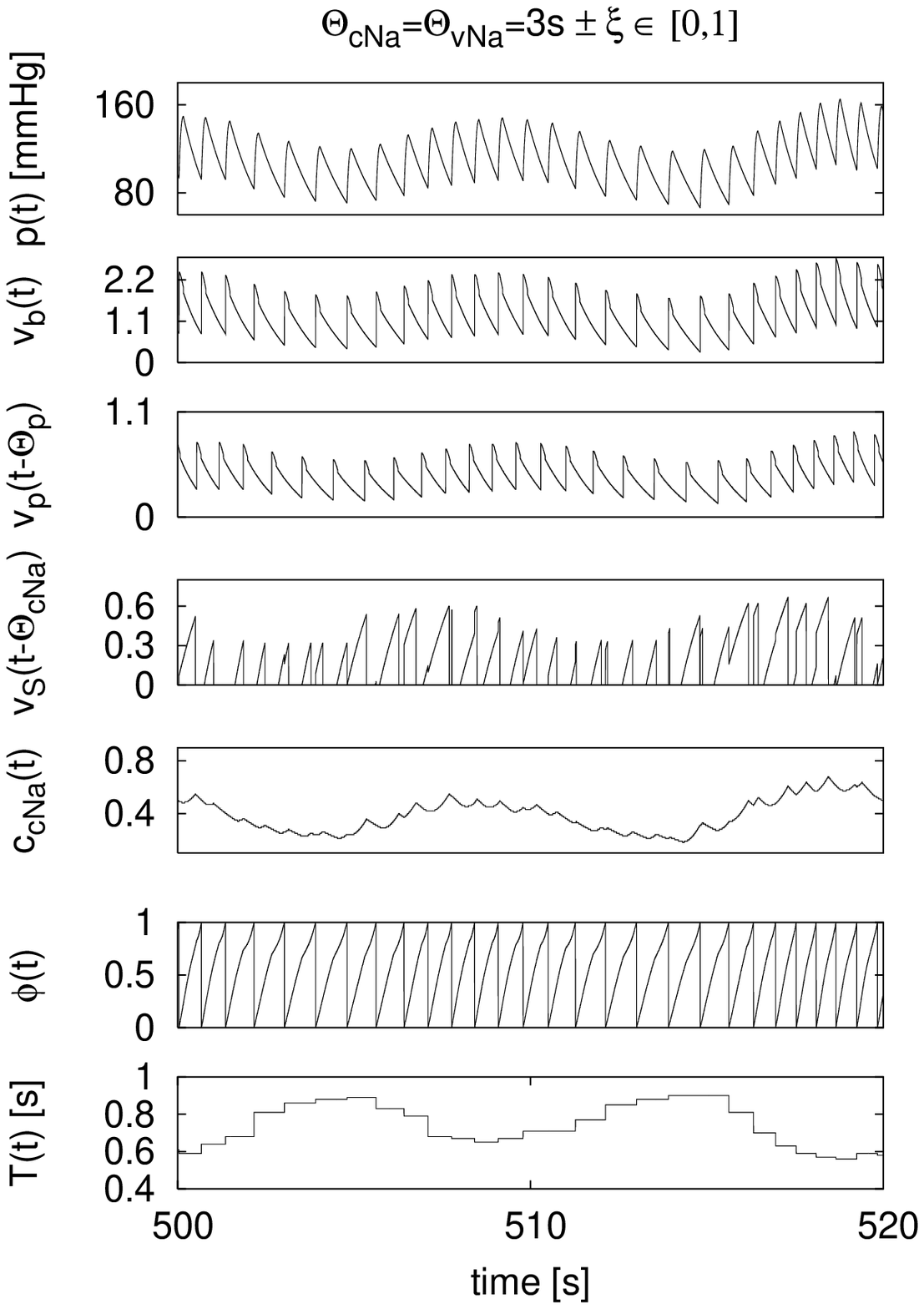}\\ [0.4cm]
\end{array}$
\end{center}

\caption{\label{simb}Time series obtained by simulating the model \textbf{(a)} without and \textbf{(b)} with stochastic delays - oscillating heart rate case}
\end{figure*}
}
 
\def\figureHistrr{
\begin{figure}[hbt]
  \includegraphics[width=0.45\textwidth]{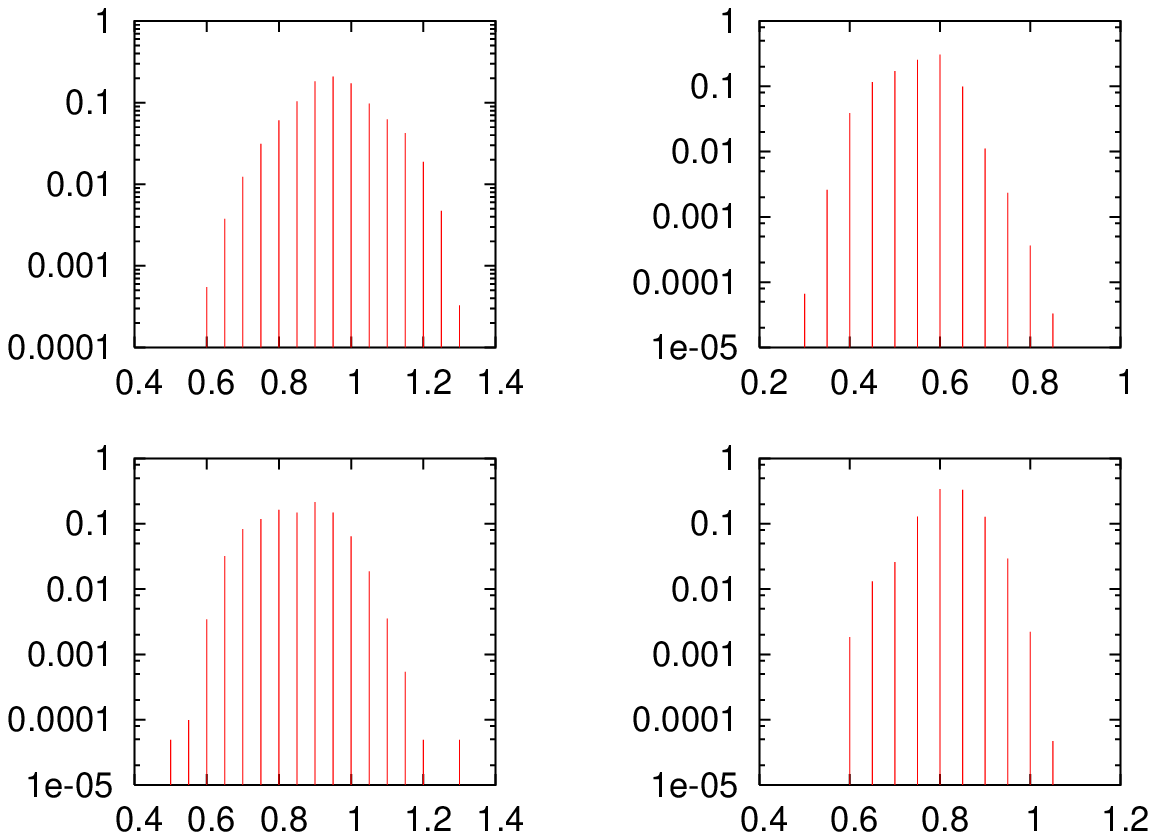}
  \caption{Histograms of time series of RR intervals from healthy subjects \cite{data}. To study the similarity to normal distribution the log scale is applied.}
  \label{histrr}
\end{figure}  
}

\def\figureHista{
\begin{figure}
\includegraphics[width=0.45\textwidth]{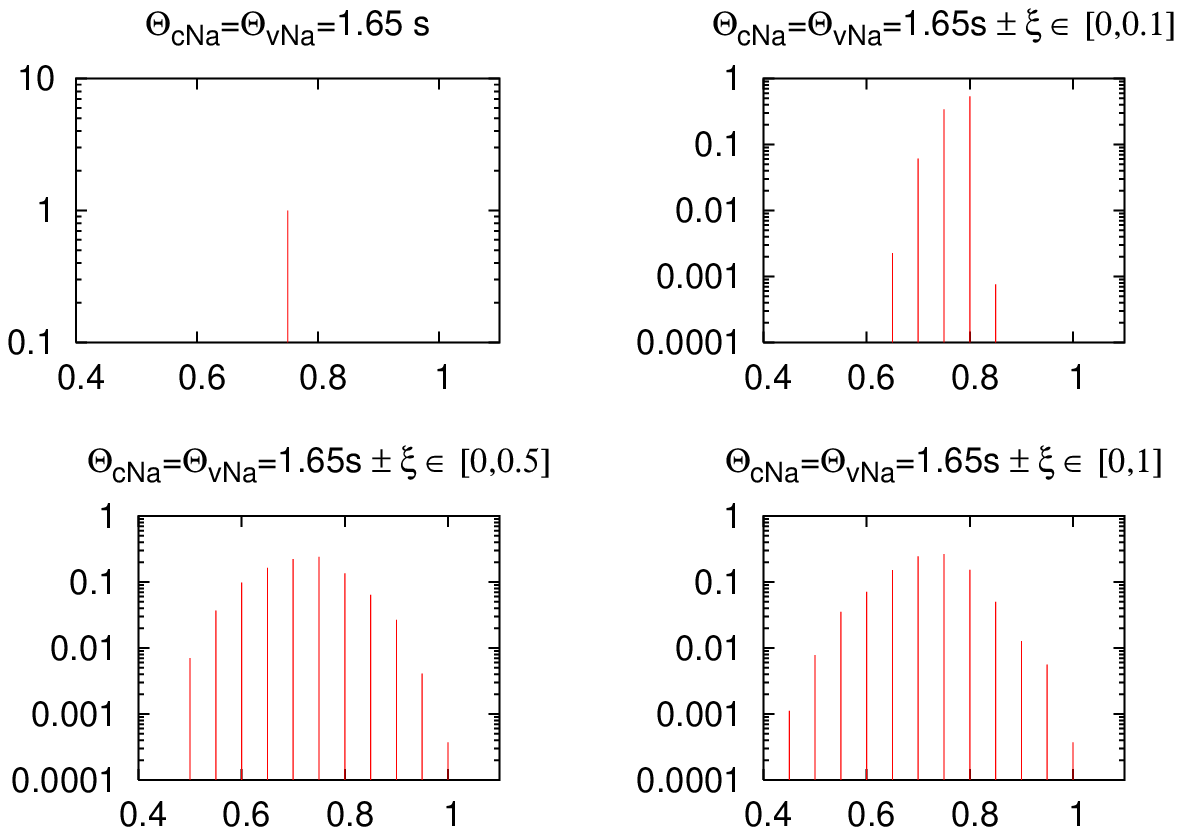}
  \caption{Histograms of heart period T - \emph{regular} heart rate case. The unperturbed evolution (top-left panel) and evolution with increasing level of noise. To study the similarity to normal distribution the log scale is applied.}
  \label{hista}
\end{figure}  
}

\def\figureHistd{
\begin{figure}
\includegraphics[width=0.45\textwidth]{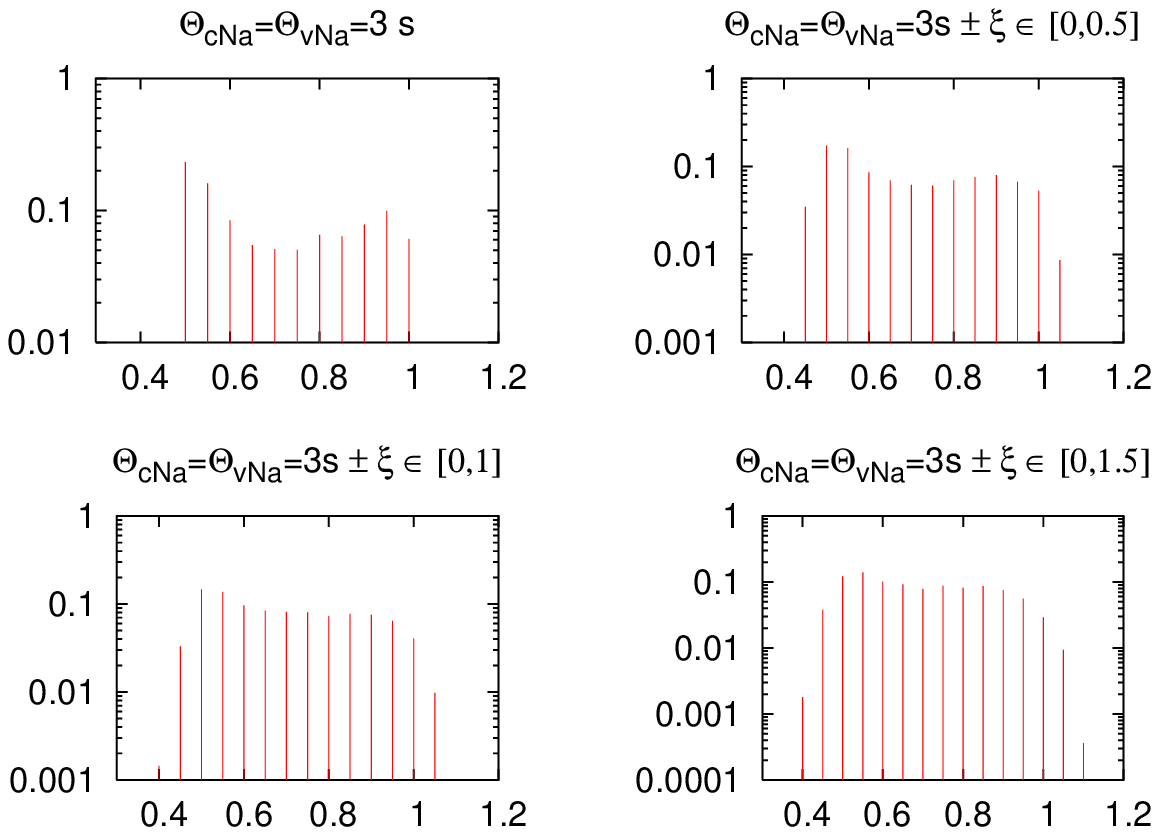}
  \caption{Histograms of heart period T - \emph{irregular} heart rate case. The unperturbed evolution (top-left panel) and evolution with increasing level of noise. To study the similarity to normal distribution the log scale is applied.}
  \label{histd}
\end{figure}  
}

\title{Influence of stochastic delays in \\Seidel-Herzel model of human cardiorespiratory system}
\date{\today}
\author{Aleksandra Dudkowska}
\email[e-mail adress:  ]{ola@iftia9.univ.gda.pl}
\author{Danuta Makowiec}
\affiliation{Institute of Theoretical Physics and Astrophysics,University of Gda\'nsk ul. Wita Stwosza 57, 80-952 Gda\'nsk, Poland}

\begin{abstract}
Experiments that discuss influence of noise to H. Seidel and H. Herzel dynamics model of human cardiovascular system are presented. Noise is introduced by considering stochastic delays in response to the sympathetic system. It appears that in the presence of the noise 10 s heart rate oscillations connected with Mayer waves are preserved. Moreover the heart rate becomes approximately normally distributed (even in unstable phase of original Seidel Herzel model), similarly like the real RR intervals data.
\end{abstract}
\pacs{87.19.Hh, 87.80.Vt, 89.75.Fb, 05.45.Pq}
\maketitle

%PACS numbers: 87.19.Hh, 87.80.Vt, 89.75.Fb, 05.45.Pq
%\frontmatter
\section{Introduction}
The human cardiovascular system is controlled by several neural and hormonal mechanisms. Among these mechanisms the autonomic nervous system via baroreceptors activity provides blood pressure control \cite{malik}. This control is realized by two parts of the autonomic system acting antagonistically: the sympathetic subsystem which activity accelerates the heart rate and the parasympathetic subsystem which activity elongates the interbeat intervals.
The finite propagation speed of neural impulses induce several time delays. Time delays are a very common feature of physiological systems \cite{mackey}. Their incorporation is essential for a realistic modelling.

Modelling of the baroreceptors-cardiac reflex dynamics has  a long history, see e.g. \cite{deboer,ottesen,eyal,Kotani,Struzik,fowl}.
One of the still widely investigated models was proposed by DeBoer {\it et al.} \cite{deboer}. In this model blood pressures, respiration, peripheral resistance and cardiac interbeat intervals are represented by a simple beat-to-beat relations.  Ottesen \cite{ottesen} modelled chronotropic (heart rate regulation) and inotropic (contractility of ventricle regulation) parts of the baroreflex-feedback mechanism. This proposition based on expanded Windkessel model, describes both sympathetic and parasympathetic nervous system, with one sympathetic time delay. 
%Stability of the model with attention to the effect of time delay is analyzed. 
The system can switch between being stable and oscillatory, and vice versa, with changes of sympathetic time delay.
Eyal { \it et al.} \cite{eyal} studied simplified linearized version of DeBoer model. They found that increase of sympathetic gain or decrease of vagal gain parameters leads to Hopf bifurcation and sustained oscillations.

The  idea of  Seidel and Herzel proposition, called SH model, consists in  entering the internal  dynamics of one heartbeat \cite{sei}. It is achieved by introducing a phase of a sinus node (the first pacemaker of the heart) and  then considering the sensitivity of the sinus node phase change to autonomic neurotransmitter kinetics and vascular dynamics. Moreover, the SH model is strongly physiologically grounded. Parameters of the model take experimentally known values and the solution restores the rhythms of heartbeat with a period of about 1 second, respiration with a period of 4 seconds and Mayer waves with a period about 10 seconds. Therefore this model is often used in modelling  other phenomena related to heart dynamics, e.g.,  synchronization between respiration and heart  rate \cite{Kotani} or  presence of 1/f fluctuations in interbeat interval signals  \cite{Struzik}.

Since interactions in SH model involve nonlinear dependencies then chaotic aspects of dynamics are present. An increase of the delays in conveying of sympathetic nervous system signals leads via Hopf bifurcation to heart rate oscillations \cite{sh2}. Such altered dynamics as the result of bifurcation is present in many mathematical models of physiological systems. Especially systems with feedbacks and time delays can exhibit both oscillatory and chaotic dynamics \cite{glass}. Bifurcation with time delays as bifurcation parameters is observed also in other models of baroreflex \cite{ottesen,eyal,magosso}.

It is known that cardiovascular signals behave deterministically, but noises of various origin are also present in the system\cite{stef}. For example, the basic neurotransmitter of the sympathetic nervous system - noradrenaline, is distributed to the heart and vessels via diffusion. Therefore this process should be represented by slow stochastically  dependent dynamics. It would be  interesting to change  SH model by considering stochastically driven  parameters describing delays in signal transfers. The goal of our investigation is to examine the stability of the solution of SH model when the delay parameters are stochastic variables. 

We focus our interest on the bifurcation regime. This regime demands careful dealing with model parameters. We have found that there is a parameter for which the solution of the original SH model has no physiological meaning. This parameter is involved in the construction of the sigmoidal function which is responsible for Windkessel time constant dependence on large vascular concentration of noradrenaline. To keep the solution within physiologically reasonable limits we propose to change a value of this parameter. The change proposed by us does not influence the model solution in the regular regime. However it allows to find solutions which are numerically stable and physiologically  rational for wide range of time delays.

For the reasons described above we present the basic model and our modification in details, Section 2, and then discuss thoroughly obtained solution, Section 3. Section 4 contains simulations when delays are stochastically changed.

\section{SH model}

\figureDiagram

SH model provides relations between different mechanisms involved in the baroreceptor-cardiac reflex in the set of the following equations: (see Fig.\ref{SH_scheme} for the visualization of  relations between variables)
\begin{itemize}
\item[---]Baroreceptor activity $\nu_{b}$:

\begin{equation}
\nu_{b}=k_{1}(p-p^{(0)})+k_{2}\frac{dp}{dt}
\label{eq1}
\end{equation} 
($p$ - blood pressure; $k_{1}=0.02 mm Hg^{-1}, k_{2}=0.00125 s\ mm Hg^{-1}, p^{(0)}=50 mmHg$ - parameters)

\item[---] Sympathetic activity $\nu_{s}$:
\begin{equation}
\nu_{s}=\max(0,\nu_{s}^{(0)}-k_{s}^{b}\nu_{b}+k_{s}^{r}|\sin(\pi f_{r}t+\Delta\phi_{s}^{r})|)
\label{eq2}
\end{equation} 
($f_{r}=0.2 s^{-1}$ - respiratory frequency; $\nu_{s}^{(0)}=0.8, k_{s}^{b}=0.7,k_{s}^{r}=0.1, \Delta\phi_{s}^{r}=0.0$ - parameters)

\item[---] Parasympathetic activity $\nu_{p}$:
\begin{equation}
\nu_{p}=\max(0,\nu_{p}^{(0)}+k_{p}^{b}\nu_{b}+k_{p}^{r}|\sin(\pi f_{r}t+\Delta\phi_{p}^{r})|)
\label{eq3}
\end{equation} 
($f_{r}$ - respiratory frequency; $\nu_{p}^{(0)}=0.0, k_{p}^{b}=0.3,k_{p}^{r}=0.1, \Delta\phi_{p}^{r}=0.0$ - parameters)

\item[---] Cardiac concentration of sympathetic transmitter $c_{cNa}$:
\begin{equation}
\frac{dc_{cNa}}{dt}=-\frac{c_{cNa}}{\tau_{cNa}}+k_{c_{cNa}}^{s}\nu_{s}(t-\theta_{cNa})
\label{dccna}
\end{equation} 
($\tau_{cNa}=2.0 s, k_{c_{cNa}}^{s}=1.2$ - parameters, $\theta_{cNa}$ - time delay)

\item[---] Sinus node phase $\varphi$:
\begin{equation}
\frac{d\varphi}{dt}=\frac{1}{T^{(0)}}f_{s}f_{p}
\end{equation} 
($f_{s}, f_{p}$ - sympathetic and parasympathetic influences; $T^{(0)}=1.1 s$ - parameter)

\item[---] Sympathetic influence on the phase velocity of the sinus node:

\begin{eqnarray}
f{s}=1+k_{\varphi}^{cNa}[&(c_{cNa}&+\nonumber\\
&(\widehat{c}_{cNa}&-c_{cNa})\frac{c_{cNa}^{n_{cNa}}}{(\widehat{c}_{cNa})^{n_{cNa}}+c_{cNa}^{n_{cNa}}}]
\end{eqnarray}

($k_{\varphi}^{cNa}=1.6, \widehat{c}_{cNa}=2.0, n_{cNa}=2.0$ - parameters)

\item[---] Parasympathetic influence on the phase velocity of the sinus node:

\begin{eqnarray}
f_{p}=1-k_{\varphi}^p[&\nu_{p,\theta_{p}}&+\nonumber\\
&(\widehat{\nu}_{p}&-\nu_{p,\theta_{p}})   \frac{\nu_{p,\theta_{p}}^{n_{p}}} {\widehat{\nu}_{p}^{n_{p}} + \nu_{p,\theta_{p}} ^{n_{p}}}]F(\varphi)
\end{eqnarray}
($\nu_{p,\theta_{p}}=\nu_{p}(t-\theta_{p})$ - delayed parasympathetic activity, $F(\varphi)$ - phase effectiveness curve, $k_{\varphi}^p=5.8, \widehat{\nu}_{p}=2.5, n_{p}=2.0$- parameters)

\item[---] Phase effectiveness curve F:
\begin{equation}
F(\varphi)=\varphi^{1.3} (\varphi-0.45)\frac{(1-\varphi)^{3}}{(1-0.8)^{3}+(1-\varphi)^3}
\end{equation}

\item[---] Cardiac contractility without saturation $S'_{i}$:
\begin{equation}
S'_{i}=S^{(0)}+k_{S}^{c}c_{cNa}+k_{S}^{t}T_{i-1}
\end{equation}
($T_{i-1}$ - duration of previous heart period; $S^{(0)}=25 mmHg, k_{S}^{c}=40 mmHg, k_{S}^{t}=10 mmHg s^{-1}$ - parameters)

\item[---] Cardiac contractility with saturation $S_{i}$:
\begin{equation}
S_{i}=S'_{i}+(\widehat{S}-S'_{i})\frac{S'^{n_{S}}_{i}}{S'^{n_{S}}_{i}+\widehat{S}^{n_{S}}}
\end{equation}
($\widehat{S}=70 mmHg, n_{S}=2.5$ - parameters)

\end{itemize}

\figureTau

\begin{itemize}

\item[---] Windkessel time 'constant' $\tau_{v}$:

\begin{eqnarray}
\tau_{v}=\tau_{v}^{(0)}-\overline{\tau_{v}} [&c_{vNa}& +\nonumber\\
&(\widehat{c}_{vNa}&-c_{vNa})\frac{c_{vNa}^{n_{vNa}}}{\widehat{c}_{vNa}^{n_{vNa}} + c_{vNa}^{n_{vNa}}}]
\label{tauv} 
\end{eqnarray}
($\tau_{v}^{(0)}=2.2 s, \overline{\tau_{v}}=1.2 s, \widehat{c}_{vNa}=10.0, n_{vNa}=1.5$ - parameters)\newline
{\it Model modification:}  Fig. \ref{figtauv} presents the dependence  of $\tau_{v}$ - Windkessel time 'constant' on $c_{vNa}$, following Eq. \ref{tauv}.
Such  dependence admits $\tau_{v}$ to be negative. However since $\tau_{v}$ is responsible for the blood pressure decay in aorta during diastolic part of the heartbeat cycle, see eq.(\ref{eqlast}), hence it must be positive.
Therefore we propose the following change in the sigmoidal function parameter: let $\widehat{c}_{vNa}=1.0$ in place of $\widehat{c}_{vNa}=10.0$.
This change becomes important only in a case when delays $\theta_{cNa}$, $\theta_{cNa}$ are greater than 3.0.

\item[---] Vascular concentration of sympathetic transmitter $c_{vNa}$:
\begin{equation}
\frac{dc_{vNa}}{dt}=-\frac{c_{vNa}}{\tau_{vNa}}+k_{c_{vNa}}^{s}\nu_{s}(t-\theta_{vNa})
\label{dcvna}
\end{equation} 
($\tau_{vNa}=2.0 s, k_{c_{vNa}}^{s}=1.2$ - parameters, $\theta_{vNa}$ - time delay)

\item[---] Blood pressure during the systolic part of the heart cycle:
\begin{equation}
p=d_{i-1}+S_{i}\frac{t-t_{i}}{\tau_{sys}} exp\left\lbrace 1-\frac{t-t_{i}}{\tau_{sys}} \right\rbrace 
\end{equation} 
($d_{i-1}$ - diastolic pressure during the systolic part of the heart cycle, $t_{i}$ - time of last contraction onset, $\tau_{sys}=0.125 s$ - parameter)

\item[---] Blood pressure during the diastolic part of the heart cycle:
\begin{equation}
\frac{dp}{dt}=-\frac{p}{\tau_{v}(t)}
\label{eqlast}
\end{equation}

\end{itemize}

\section{Numerical simulation of the model}
The adapted Runge-Kutta method of fourth order with a constant step size (h=0.001s) \cite{num} was used. The ring buffers are introduced to store history of sympathetic and vagal activities. The first 500 seconds of evolution are neglected, as the transient time, and then the results are collected and analyzed. The results with parameters setting as in the original Seidel-Herzel proposition are presented in Fig. \ref{ts_norand}. Subsequently, starting from the top of Fig. \ref{ts_norand}, we show: 
\begin{itemize}
\item The blood pressure $p$ which is characterized by rapid increase during the systole, and gentle decrease during the diastole. Minimum (diastolic - about 80 mmHg) and maximum (systolic - about 140 mmHg) values of this wave nicely reproduce physiological one's.
\item Baroreceptors activity $v_b$ that basically is proportional to the blood pressure. Hence the waves in the first and the second panels are similar to each other.
\item The parasympathetic activity $v_p$ that affects the heart with the time delay $\theta_p$ and because of this delay the oscillations of the activity are shifted in time. Since  transmission of the signal goes fast (mechanism of transmission is different from diffusion) then the delay can be assumed as constant and relatively small, i.e., $\theta_{p} = 0.5 s$.
\item The sympathetic system activity $v_s$ enters into the model with two delays: $\theta_{cNa}$ and $\theta_{vNa}$ corresponding to its influence to the heart and vascular system, respectively. Delays $\theta_{cNa}$, $\theta_{vNa}$ represent times needed by transmitters to transfer the information by diffusion. The activity  affecting the heart $v_s(t-\theta_{cNA})$ is presented in the figure.
\item Cardiac $c_{cNa}$ and vascular $c_{vNa}$ noradrenaline concentrations which take the same values.
\item The phase of the sinus node $\varphi$ which mimics heart beating (integrate-and-fire-model) and establishes the rhythm of the whole system. All variables of the model oscillate with a period about 0.8 s.
\item The sequence of heart periods $T_i$ that represents the time interval between two consecutive heart beats.
\end{itemize} 

Our results are slightly different from results presented in \cite{sei}. To explain  let us observe that there is some inconsistency in the results of \cite{sei}. For example, it is easy to estimate (directly from equations (\ref{eq2}) and (\ref{eq3})) that the maximum and minimum values of $v_p$ and $v_s$ plotted in Fig. 4 of \cite{sei} should be different: $v_p$ should be smaller and maxima of $v_s$ should be greater than it is shown.

\figureSimulation

Sympathetic and parasympathetic activities are influenced by respiratory neurons. This influence is described by the sine function in Eq.(\ref{eq2}),(\ref{eq3}). Blood pressure and heart period variations caused by respiratory are called respiratory sinus arrhythmia (RSA). Respiratory modulation of neural activities is clearly visible in Fig.\ref{ts_norand}(a). All waveforms are modulated with the frequency of respiration. As in our further simulations we concentrate on mechanisms other than respiration then the sine function is replaced by the average value, and the respiratory modulation of neural activities is neglected. Fig. \ref{ts_norand}(b) shows the solution of the model with the neglected respiratory modulation. In the absence of RSA the heart period is approximately constant but it is not settled to a one point. Let us call this solution a regular heart rate.

We present two representative examples to illustrate how time delays influence the solution of the model. The first bifurcation diagram is obtained for $\theta_{vNa} = 1.65s$ and increasing $\theta_{cNa}$, see Fig. \ref{figbifc165}, and the second one is for $\theta_{vNa} = 3 s$ and increasing $\theta_{cNa}$,  Fig.\ref{figbifc3}. One can easily separate regions of the regular heart rate ($\theta_{vNa}=1.65$: $\theta_{cNa} < ~2 s$ and $\theta_{vNa}=3$: $\theta_{cNa} < ~ 0.6s$) from the heart rate with oscillations ($\theta_{vNa}=1.65$: $~2<\theta_{cNa}<3.5 s$  and  $\theta_{vNa}=3$: $  ~0.6<\theta_{cNa} <3.5s$). The influence of delays is not  symmetric:  the setting $\theta_{vNa}=1.65 ,\ \theta_{cNa}=3.0 $ leads to heart periods $0.45<T<1.1$ while the setting $\theta_{vNa}=3. ,\ \theta_{cNa}=1.65 $ leads to $0.55<T<0.9$.

\figureBifcrega

\figureBifcregb

\section{Stochastic approach}
Let us assume that  a time, needed for a signal to transfer information  from the brain through the sympathetic nerves  to the heart, varies stochastically around some mean value. To incorporate this assumption let us introduce  random variables in place of fixed delays in the activity function of the  sympathetic nervous system. 
So Eq. (\ref{dccna}) and  (\ref{dcvna}) obtain the following form:
\begin{equation}
\frac{dc*}{dt}=-\frac{c*}{\tau*}+k_{*}^{s}\nu_{s}(t-(\theta_{*}+\xi_{*}))
\end{equation}
where $\xi_{*}$ are random variables with uniform distribution and $\xi_{*} \in [-\widehat{\xi_{*}},\widehat{\xi_{*}}]$, and * represents either cNa cardiac or vNa vascular noradrenaline concentration. The random numbers are generated by standard 48-bit arithmetic functions accessible in C compilers in Unix systems. 
Noradrenaline needs to be removed from the synaptic gap and broken down at another location. This process takes much longer time than a time step of model simulation. Noradrenaline concentration acts like a buffer so the distribution of sympathetic impulses over a heart cycle is not very important \cite{sh2}. Therefore we assign  random  values  to  delays for  each cardiac cycle separately. 
One can model the  noise $\xi_{*}$ by  a continuous variable also.  In this case, at  each numerical step  a different random  value to the delay is  set. The randomness put into the system in such a way does not cause significant  change in solutions of the model.

\figureHistrr

To avoid effects related to respiration we again consider the system with constant respiration only. The influence of the stochastic perturbation to the system will be examined when the heart beats regularly: $\theta_{cNa} = \theta_{vNa} = 1.65 s$, and when the heart rate oscillates: $\theta_{cNa} = \theta_{vNa} = 3 s$.
Statistical properties of the heart rate will be presented by histograms - the probability that a given period is present in a series. For comparison, let us show  few examples of histograms made from real heart rate series (so-called normal sinus rhythm series),  Fig. \ref{histrr}. Although each of the histogram is different from others but the parabola-like shape in log-scale suggests that the Gaussian distribution of heart rate intervals can be used as the first approximation to the histograms.

\subsection{Stochastic perturbation to regular heart rating}

\figureSimulationa

If the stochastic time delays are considered then the sympathetic activity becomes irregular what affects others variables --- the heart rate also. Fig. \ref{sima} shows the time series obtained when the stochastic delays are applied. The regular waveforms received in deterministic case for the sympathetic activity, compare to Fig. \ref{ts_norand}b,  now, at the presence of stochastic delay, are varying  irregularly. The unperturbed evolution provides the histogram with one value of heart period only, see Fig. \ref{hista}.  Depending on the magnitude of added noise this line is broadened. Notice, that at high level of the noise, namely if $ 0.65 s <\theta_{cNa}, \theta_{vNa} < 2.65s$, the histogram shape  reminds distributions of real series. 

\figureHista

\subsection{Stochastic perturbation to irregular heart rating}

\figureSimulationb

Fig. \ref{simb}a presents unperturbed time series for $\theta_{cNa} = \theta_{vNa}=3 $ while Fig. \ref{simb}b shows series when delays are $ 2<\theta_{cNa}, \theta_{vNa} <4$  taken at random.

The SH model is able to generate sustained 10 s rhythms in the bifurcation regime \cite{sei} which can be related to Mayer waves known from physiology. However the statistical properties of heart period are strange.
The histogram of the heart beat intervals in case of the unperturbed dynamics takes the form of a double peaked distribution with sharp decay of the left and right wings, Fig. \ref{histd}. The shortest and the longest periods appear with highest probability. The heart period looks like switching between these two values. Such a distribution is not resembling any distribution found for real RR time series.  
\figureHistd

Stochastic noise influences the heart rate but the main oscillations of the deterministic solution such as Mayer waves are partially preserved, see Fig.\ref{simb}(b). However, the distribution of the intervals is wider --- the intervals from peaks  are less probable. When the noise level is such that conditions for a regular heart rating emerge, i.e.,  $ 1.5 <\theta_{cNa}, \theta_{vNa} <4.5$, then the distribution almost restores the normal shape, see Fig. \ref{histd} right-bottom plot.

\section{Conclusion}
In this paper we investigated the dynamic properties of stochastic nonlinear model of the cardiovascular system. The stochastic model was obtained by allowing random changes to bifurcation parameters of the Seidel-Herzel model. Using the computer simulations the role of the stochastic perturbations was studied. It is shown, that  stochastic delays lead to time series with probabilistic properties  similar to those obtained from real time series for healthy humans. Our results ensure that incorporating with random variables and in consequence considering stochastic differential equations in place of deterministic ones, we gain more realistic description of the cardiovascular system.
Moreover, the introduced change in sigmoidal function parameter allows to explore the model for wide range of time delays. As SH model is widely used  to reproduce key features of human heartbeat dynamics (e.g.\cite{Kotani,Struzik}) the possibility of carrying out simulation for wide range of parameters is very important for further study. 

{\noindent\bf Acknowledgments}
We wish to acknowledge the support of the Rektor of Gda\'nsk University --- project: BW $\slash$5400-5-0166-5

\end{document}